\documentclass[10pt,a4paper]{article}
\usepackage{amsmath}
\usepackage{amssymb}
\usepackage{amsfonts}
\usepackage{amstext}
\usepackage{amsbsy}
\usepackage[mathscr]{eucal}
\usepackage{graphicx}
\newcommand{\beq}{\begin{equation}}
\newcommand{\eeq}{\end{equation}}
\newcommand{\beqa}{\begin{eqnarray}}
\newcommand{\eeqa}{\end{eqnarray}}
\newcommand{\ket}[1]{| #1 \rangle}
\newcommand{\bra}[1]{\langle #1 |}

\makeindex
\title{\Large\textbf{Quantum entanglement measure based on wedge product}}

\author{\textit{ Hoshang Heydari}\\
        \small\textit{Institute of Quantum
Science, Nihon University,}\\
\small\textit{1-8 Kanda-Surugadai, Chiyoda-ku, Tokyo 101-8308, Japan
}}

\date{}
\begin{document}

\maketitle \thispagestyle{empty}

\maketitle
\begin{abstract}
We construct an entanglement measure that coincides with the
generalized concurrence for a general pure bipartite state based
on wedge product. Moreover, we construct an entanglement measure
for pure multi-qubit states, which are entanglement monotone.
Furthermore, we generalize our result on a general pure
multipartite state.
\end{abstract}

\section{Introduction}
Quantum entanglement is one of the most interesting properties of
quantum mechanics. It has become an essential resource for quantum
information (including quantum communication and quantum computing)
developed in recent years, with some potential applications such as
quantum cryptography \cite{Bennett84,Ekert91}and quantum
teleportation \cite{Bennett93}.  Quantification of a multipartite
state entanglement \cite{Lewen00,Dur99} is quite difficult and is
directly linked to  algebra, geometry, and functional analysis. The
definition of separability and entanglement of a multipartite state
was introduced in\cite{Vedral97}, following the definition of
bipartite states, given by Werner \cite{Werner89}. One of the widely
used measures of entanglement of a pair of qubits is entanglement
 of formation and the concurrence, that gives an analytic formula for the entanglement of formation
 \cite{Bennett96,Wootters98,Wootters00}.
In recent years, there have been some proposals to generalize this
measure on general pure bipartite states
\cite{Uhlmann00,Audenaert,Rungta01,Gerjuoy} and on multipartite
states
 \cite{Albeverio,Bhaktavatsala,Akhtarshenas,Hosh2,Hosh3}.  There also have been
many works on entanglement measures which rely on a wedge product,
e.g., through the definition of hyper-determinants. For example, F.
Verstraete \emph{et al}. \cite{Vers1} have considered a single copy
of a pure four-partite state of qubits and investigated its behavior
under stochastic local quantum operation and classical
communication(SLOCC)\cite{Dur00}, which gave a classification of all
different classes of pure states of four qubits. They have also
shown that there exist nine families of states corresponding to nine
different ways of entangling four qubits. Note that, all homogeneous
positive functions of pure states that are invariant under
 SLOCC operations are called entanglement monotones.
F. Verstraete \emph{et al}. \cite{Vers2} have also presented a
general mathematical framework to describe local equivalence classes
of multipartite quantum states under the action of local unitary and
local filtering operations. Their analysis has lead to the
introduction of entanglement measures for the multipartite states,
and the optimal local filtering operations maximizing these
entanglement monotones were obtained.
E. Briand, \cite{Briand1} \emph{et.} al have obtained a complete and
minimal set of 170 generators for the algebra of
$SL(2,\mathbf{C})^{\times 4}$-covariants of a binary quadrilinear
form. Interpreted in terms of a four qubit system, this describes in
particular the algebraic varieties formed by the orbits of local
filtering operations in its projective Hilbert space.
E. Briand, \cite{Briand2} \emph{et.} al have also studied  the
invariant theory of trilinear forms over a three-dimensional complex
vector space, and apply it to investigate the behavior of pure
entangled three-partite qutrit states and their normal forms SLOCC
operations. They described the orbit space of the SLOCC group
$SL(3,\mathbf{C})^{\times 3}$ both in its affine and projective
versions in terms of a very symmetric normal form parameterized by
three complex numbers. They have also shown that the structure of
the sets of equivalent normal forms is related to the geometry of
certain regular complex polytopes.
A. Miyake and M. Wadati \cite{Miyake} have  explored quantum search
from the geometric viewpoint of a complex projective space. They
have shown that the optimal quantum search can be geometrically
identified with the shortest path along the geodesic joining a
target state, an element of the computational basis, and such an
initial state as overlaps equally, up to phases, with all the
elements of the computational basis. They have also calculated the
entanglement through the algorithm for any number of qubits $n$ as
the minimum Fubini-Study distance to the submanifold formed by
separable states in Segre embedding, and find that entanglement is
used almost maximally for large $n$.
Recently, P\'{e}ter L\'{e}vay \cite{Levay1} have constructed a class
of multi-qubit entanglement monotones which was based on
construction of C. Emary \cite{Emary}. His construction is based on
bipartite partitions of the Hilbert space and the invariants are
expressed in terms of the Pl\"{u}cker coordinates of the
Grassmannian.
We have also constructed entanglement monotones for multi-qubit
states based on Pl\"{u}cker coordinate equations of Grassmann
variety, which are central notion in geometric invariant theory
\cite{Hosh4}. However, we do have different approaches and
construction to solve the problem of quantifying multipartite states
compare to those of L\'{e}vay and Emary. In this paper, we will
construct a measure of entanglement by using the algebraic
definition of wedge product without touching the geometrical
structure of algebraic projective variety defining this measure of
entanglement. In particular, in section \ref{Mgbipartite} we will
derive a measure of entanglement that coincides with
 the generalized concurrence for a general pure
bipartite state, based on an algebraic point of view, using wedge
product, a useful tool from multi-linear algebra, which is mostly
used in algebraic and differential geometry and topology, in
relation with differential forms. In section \ref{Mqubit}, we will
construct
 entanglement monotones for a general pure multi-qubit state.
Moreover, we generalize our construction on a general pure
multipartite state in section \ref{Mpartite}. To make this note
self-contained in section \ref{Mlinear}, we will give a preview
introduction to multi-linear algebra.
 Let us denote a general,
multipartite quantum system with $m$ subsystems by
$\mathcal{Q}=\mathcal{Q}_{m}(N_{1},N_{2},\ldots,N_{m})$ $
=\mathcal{Q}_{1}\mathcal{Q}_{2}\cdots\mathcal{Q}_{m}$, consisting of
a state $
\ket{\Psi}=\sum^{N_{1}}_{k_{1}=1}\cdots\sum^{N_{m}}_{k_{m}=1}
\alpha_{k_{1},\ldots,k_{m}} \ket{k_{1},\ldots,k_{m}} $ and, let
$\rho_{\mathcal{Q}}=\sum^{\mathrm{N}}_{n=1}p_{n}\ket{\Psi_{n}}\bra{\Psi_{n}}$,
for all $0\leq p_{n}\leq 1$ and $\sum^{\mathrm{N}}_{n=1}p_{n}=1$,
denote a density operator acting on the Hilbert space $
\mathcal{H}_{\mathcal{Q}}=\mathcal{H}_{\mathcal{Q}_{1}}\otimes
\mathcal{H}_{\mathcal{Q}_{2}}\otimes\cdots\otimes\mathcal{H}_{\mathcal{Q}_{m}},
$ where the dimension of the $j$th Hilbert space is given  by
$N_{j}=\dim(\mathcal{H}_{\mathcal{Q}_{j}})$. We are going to use
this notation throughout this paper, i.e., we denote a mixed pair of
qubits by $\mathcal{Q}_{2}(2,2)$. The density operator
$\rho_{\mathcal{Q}}$ is said to be fully separable, which we will
denote by $\rho^{sep}_{\mathcal{Q}}$, with respect to the Hilbert
space decomposition, if it can  be written as $
\rho^{sep}_{\mathcal{Q}}=\sum^\mathrm{N}_{n=1}p_{n}
\bigotimes^m_{j=1}\rho^{n}_{\mathcal{Q}_{j}}$,
$\sum^\mathrm{N}_{n=1}p_{n}=1 $, for some positive integer
$\mathrm{N}$, where $p_{n}$ are positive real numbers and
$\rho^{n}_{\mathcal{Q}_{j}}$ denotes a density operator on the
Hilbert space $\mathcal{H}_{\mathcal{Q}_{j}}$. If
$\rho^{p}_{\mathcal{Q}}$ represents a pure state, then the quantum
system is fully separable if $\rho^{p}_{\mathcal{Q}}$ can be written
as
$\rho^{sep}_{\mathcal{Q}}=\bigotimes^m_{j=1}\rho_{\mathcal{Q}_{j}}$,
where $\rho_{\mathcal{Q}_{j}}$ is a density operator on
$\mathcal{H}_{\mathcal{Q}_{j}}$. If a state is not separable, then
it is called an entangled state. Some of the generic entangled
states are called Bell states and $\mathrm{EPR}$ states.


\section{Multilinear algebra}\label{Mlinear}
In this section, we review the definitions and properties of
multilinear algebra and exterior algebra. Multilinear algebra
extends the methods of linear algebra, which builds on the concept
of tensor. The exterior algebra $\bigwedge(V)$ or Grassmann algebra
of a given vector space $V$ is a certain unital associative algebra,
which contains $V$ as a subspace and its multiplication, known as
the wedge product or the exterior product
 written as $\wedge$.  The wedge product is associative and bilinear.
Now, let us consider the complex vector spaces
$\mathrm{V}_{1},\mathrm{V}_{2},\ldots, \mathrm{V}_{m}$ to be
vector spaces, where $\dim(\mathrm{V}_{j})=N_{j},  ~ \forall
j=1,2,\ldots, m $. Then, we define a tensor of type $(m,n)$ on
$\mathrm{V}_{1},\mathrm{V}_{2},\ldots, \mathrm{V}_{m}$ as follows
\begin{eqnarray}
\mathcal{T}^{m}_{n}(\mathrm{V}_{1},\mathrm{V}_{2},\ldots,
\mathrm{V}_{m})&=&
  \mathcal{L}(\mathrm{V}_{1},\mathrm{V}_{2},\ldots,\mathrm{V}_{m};
  \mathrm{V}^{*}_{1},\mathrm{V}^{*}_{2},\ldots,\mathrm{V}^{*}_{n})\\\nonumber&=&
\mathrm{V}_{1}\otimes\mathrm{V}_{2}\otimes\cdots\otimes\mathrm{V}_{m}
\otimes\mathrm{V}^{*}_{1}\otimes\mathrm{V}^{*}_{2}\otimes\cdots\otimes\mathrm{V}^{*}_{n},
\end{eqnarray}
where $\mathrm{V}^{*}_{j}=\mathcal{L}(\mathrm{V}_{j};\mathbf{C})~
\forall j=1,2,\ldots, m $ is the space of linear applications
$\mathrm{V}_{j}\longrightarrow\mathbf{C}$ and is called dual of
$\mathrm{V}_{j}$. For any basis $e_{i}$ of $\mathrm{V}_{j}$ and
$e^{j}$ the dual basis of $e_{i}$ defined by
$e^{j}(e_{i})=\delta^{j}_{i}$, we have the following linear representation
\begin{eqnarray}
T=T^{i_{1},i_{2},\ldots,i_{m}}_{j_{1},j_{2},\ldots,j_{n}}
e_{i_{1}}\otimes e_{i_{2}}\otimes\cdots \otimes e_{i_{m}}\otimes
e^{j_{1}}\otimes e^{j_{2}}\otimes\cdots \otimes e^{j_{n}}.
\end{eqnarray}
For example, we have
$\mathcal{T}^{1}_{0}(\mathrm{V}_{1},\mathrm{V}_{2},\ldots,
\mathrm{V}_{m})=\mathrm{V}_{1}$ and
$\mathcal{T}^{0}_{1}(\mathrm{V}_{1},\mathrm{V}_{2},\ldots,
\mathrm{V}_{m})=\mathrm{V}^{*}_{1}$.  Let $S_{m}$ be a group of
permutations $(1,2,\ldots, m)$. Then, we call the tensor
$v_{1}\otimes v_{2}\otimes\cdots \otimes
v_{m}\in\mathcal{T}^{m}_{0}(\mathrm{V}_{1},\mathrm{V}_{2},\ldots,
\mathrm{V}_{m})$ symmetric if
\begin{eqnarray}
v_{1}\otimes v_{2}\otimes\cdots \otimes v_{m}=v_{\pi(1)}\otimes
v_{\pi(2)}\otimes\cdots \otimes v_{\pi(m)},
\end{eqnarray}
for all $\pi\in S_{m}$. The space of symmetric tensor is denoted by
$\mathcal{S}^{m}_{0}(\mathrm{V}_{1},\mathrm{V}_{2},\ldots,
\mathrm{V}_{m})$. Moreover, we call the tensor $v_{1}\otimes
v_{2}\otimes\cdots \otimes
v_{m}\in\mathcal{T}^{m}_{0}(\mathrm{V}_{1},\mathrm{V}_{2},\ldots,
\mathrm{V}_{m})$ skew-symmetric if
\begin{eqnarray}
v_{1}\otimes v_{2}\otimes\cdots \otimes
v_{m}=\epsilon(\pi)v_{\pi(1)}\otimes v_{\pi(2)}\otimes\cdots
\otimes v_{\pi(m)},
\end{eqnarray}
for all $\pi\in S_{m}$, where $\epsilon(\pi)$ is the signature of
permutation $\pi$. The space of skew-symmetric tensor is denoted
by $\Lambda^{m}_{0}(\mathrm{V}_{1},\mathrm{V}_{2},\ldots,
\mathrm{V}_{m})$. Furthermore, we have the following mapping
\begin{equation}\label{Wedge}
\begin{array}{ccc}
 \mathrm{Alt}^{m}: \mathcal{T}^{m}_{0}(\mathrm{V}_{1},\mathrm{V}_{2},\ldots,
\mathrm{V}_{m})&\longrightarrow
&\Lambda^{m}_{0}(\mathrm{V}_{1},\mathrm{V}_{2},\ldots,
\mathrm{V}_{m}) \\
  v_{1}\otimes v_{2}\otimes\cdots \otimes
v_{m}&\longmapsto & v_{1}\wedge v_{2}\wedge\cdots \wedge v_{m}
 \\
\end{array},
\end{equation}
where the alternating map is defined by
$\mathrm{Alt}^{m}(v_{1}\otimes v_{2}\otimes\cdots \otimes
v_{m})=v_{1}\wedge v_{2}\wedge\cdots \wedge
v_{m}=\frac{1}{m!}\sum_{\pi\in
S_{m}}\epsilon(\pi)v_{\pi(1)}\otimes v_{\pi(2)}\otimes\cdots
\otimes v_{\pi(m)}$. For example, for $m=2$ we have
\begin{equation}
\begin{array}{ccc}
 \mathrm{Alt}^{2}: \mathcal{T}^{2}_{0}(\mathrm{V}_{1},\mathrm{V}_{2})&\longrightarrow
&\Lambda^{2}_{0}(\mathrm{V}_{1},\mathrm{V}_{2}) \\
  v_{1}\otimes v_{2}&\longmapsto &\mathrm{Alt}^{2}(v_{1}\otimes v_{2})= v_{1}\wedge v_{2}=v_{1}\otimes
v_{2}-v_{2}\otimes v_{1}
 \\
\end{array}.
\end{equation}
The essential property of wedge product is that it is alternating
on $V$, that is $\upsilon\wedge\upsilon=0$ for all vector
$\upsilon\in V$ and
$\upsilon_{1}\wedge\upsilon_{2}\wedge\cdots\wedge\upsilon_{n}=0$,
whenever $\upsilon_{1},\upsilon_{2},\ldots,\upsilon_{n}\in V$ are
linearly dependent.

\section{Measure of entanglement for general bipartite state}
\label{Mgbipartite}
 In this section, we will directly construct a measure of entanglement for a
 general pure bipartite state $\mathcal{Q}^{p}_{2}(N_{1},N_{2})$, based on the wedge
 product. So
 let $\Lambda_{\mu,\nu}=v_{\mu}\wedge v_{\nu}$,
$v_{\mu}=(\alpha_{\mu,1},\alpha_{\mu,2},\ldots,\alpha_{\mu,N_{2}}),
~v_{\nu}=(\alpha_{\nu,1},\alpha_{\nu,2},\ldots,\alpha_{\nu,N_{2}})$
and $\overline{\Lambda}_{\mu,\nu}$ denote the complex conjugate of
$\Lambda_{\mu,\nu}$. Then, a measure of entanglement for a quantum
system $\mathcal{Q}^{p}_{2}(N_{1},N_{2})$ is given by
\begin{equation}\label{bicon}
\mathcal{E}(\mathcal{Q}^{p}_{2}(N_{1},N_{2}))=
\left(\mathcal{N}_{2}\sum^{N_{1}}_{\nu>\mu=1}\Lambda_{\mu,\nu}\overline{\Lambda}_{\mu,\nu}
\right)^{\frac{1}{2}}.
\end{equation}
Moreover, if we write the coefficients $\alpha_{i_{1},i_{2}}$, for
all $1\leq i_{1}\leq N_{1}$ and $1\leq i_{2}\leq N_{2}$ in form of
a $N_{1}\times N_{2}$ matrix as below
\begin{equation}\label{ggg}
   \mathrm{M}=\left( \begin{array}{cccc}
      \alpha_{1,1} & \alpha_{1,2} & \cdots & \alpha_{1,N_{2}} \\
      \alpha_{2,1} & \alpha_{2,2} & \cdots & \alpha_{2,N_{2}} \\
     \vdots & \vdots & \ddots & \vdots \\
      \alpha_{N_{1},1} & \alpha_{N_{1},2} & \cdots & \alpha_{N_{1},N_{2}} \\
    \end{array}\right),
\end{equation}
then
$\Lambda_{\mu,\nu}(\mathrm{M})=\upsilon_{\mu}(\mathrm{M})\wedge
v_{\nu}(\mathrm{M})$, where, the vectors
$\upsilon_{\mu}(\mathrm{M})$ and $ v_{\nu}(\mathrm{M})$ refer to
different rows of the matrix $\mathrm{M}$. As an example let us
look at the quantum system $\mathcal{Q}^{p}_{2}(2,2)$ representing
a pair of qubits. Then, the expression for a measure of
entanglement for such state using the above equation (\ref{bicon})
is given by
\begin{eqnarray}\label{pairq}
\mathcal{E}(\mathcal{Q}^{p}_{2}(2,2))&=&\left(\mathcal{N}_{2}\Lambda_{1,2}(\mathrm{M})\overline{\Lambda}_{1,2}(\mathrm{M})
\right)^{\frac{1}{2}}
=\left(2\mathcal{N}|\alpha_{1,1}\alpha_{2,2}-\alpha_{2,1}\alpha_{1,2}|^{2}
\right)^{\frac{1}{2}}
\\\nonumber&=&2|\alpha_{1,1}\alpha_{2,2}-\alpha_{2,1}\alpha_{1,2}|,
\end{eqnarray} where $\Lambda_{1,2}(\mathrm{M})$ is given by
\begin{eqnarray}\nonumber
\Lambda_{1,2}(\mathrm{M})&=&v_{1}\wedge v_{2} =
(\alpha_{1,1},\alpha_{1,2})\otimes
(\alpha_{2,1},\alpha_{2,2})-(\alpha_{2,1},\alpha_{2,2})\otimes
(\alpha_{1,1},\alpha_{1,2})
\\&=&
(0,\alpha_{1,1}\alpha_{2,2}-\alpha_{2,1}\alpha_{1,2}
,\alpha_{1,2}\alpha_{2,1}-\alpha_{2,2}\alpha_{1,1},0)
\end{eqnarray}
 for $\mathcal{N}_{2}=2$. The measure of entanglement for the general bipartite
 state defined in equation (\ref{bicon}) coincides with the generalized
 concurrence
\begin{eqnarray}
    \mathcal{C}(\mathcal{Q}^{p}_{2}(N_{1},N_{2}))&=&
\left ( 4\mathcal{N}_{2} \sum^{N_{1}}_{l_{1}>k_{1}=1}
\sum^{N_{2}}_{l_{2}>k_{2}=1}
\left|\alpha_{k_{1},k_{2}}\alpha_{l_{1},l_{2}}-\alpha_{k_{1},l_{2}}\alpha_{l_{1},k_{2}}\right|^{2}
\right )^{1/2},
\end{eqnarray}
 defined in \cite{Uhlmann00,Audenaert,Rungta01,Gerjuoy}, and in
 particular equation (\ref{pairq}), that gives the concurrence of a pair of qubits,
 first time defined in \cite{Wootters98,Wootters00}.

\section{Entanglement measure for multi-qubit
states}\label{Mqubit} In this section, we will construct a measure
of entanglement for multi-qubit states, based on exterior product.
Let the operator
$\Lambda_{1,2}(\mathrm{M}_{j})=\upsilon_{1}(\mathrm{M}_{j})\wedge
\upsilon_{2}(\mathrm{M}_{j})$ be the wedge product between row
number one and two of a given matrix $\mathrm{M}_{j}$, which is
constructed by coefficient of a quantum system
$\mathcal{Q}^{p}_{m}(2,2,\ldots,2)$, e.g., we define
\begin{equation}
\begin{array}{c}
  \mathrm{M}_{1}=\left(%
\begin{array}{cccc}
  \alpha_{1,1,\ldots,1} & \alpha_{1,1,\ldots,2} & \ldots & \alpha_{1,2,\ldots,2}\\
  \alpha_{2,1,\ldots,1} & \alpha_{2,1,\ldots,2} & \ldots & \alpha_{2,2,\ldots,2}\\
\end{array}%
\right), \\
 \mathrm{M}_{2}=\left(%
\begin{array}{cccc}
  \alpha_{1,1,\ldots,1} & \alpha_{1,1,\ldots,2} & \ldots & \alpha_{2,1,\ldots,2}\\
  \alpha_{1,2,\ldots,1} & \alpha_{1,2,\ldots,2} & \ldots & \alpha_{2,2,\ldots,2}\\
\end{array}%
\right), \\
  \vdots \\
  \mathrm{M}_{m}=\left(%
\begin{array}{cccc}
  \alpha_{1,1,\ldots,1} & \alpha_{1,1,\ldots,1} & \ldots & \alpha_{2,2,\ldots,1}\\
  \alpha_{1,1,\ldots,2} & \alpha_{1,1,\ldots,2} & \ldots & \alpha_{2,2,\ldots,2}\\
\end{array}%
\right), \\
\end{array}
\end{equation}
where $\mathrm{M}_{j}$ for $1\leq j\leq m$ are $2$-by-$2^{m-1}$
matrices and, e.g., the matrix $\mathrm{M}_{2}$ is constructed by
permutation of indices  of  the matrix $\mathrm{M}_{1}$. Then, we
can define a measure of entanglement for multi-qubit states by
\begin{eqnarray}
\mathcal{E}(\mathcal{Q}^{p}_{m}(2,2\ldots,2))&=&\left(\mathcal{N}_{m}\sum^{m}_{j=1}\Lambda_{1,2}(\mathrm{M}_{j})
\overline{\Lambda}_{1,2}(\mathrm{M}_{j})\right)^{1/2}\\\nonumber
&=&\left(\sum^{m}_{j=1}[\upsilon_{1}(\mathrm{M}_{j})\wedge
\upsilon_{2}(\mathrm{M}_{j})][\overline{\upsilon_{1}(\mathrm{M}_{j})\wedge
\upsilon_{2}(\mathrm{M}_{j})}]\right)^{1/2},
\end{eqnarray}
where $\mathcal{N}_{m}$ is a normalization constant. This measure of
entanglement is entanglement monotones. However, an algebraic proof
of this statement seems  difficult,  but this can be seen from
geometrical structure called Grassmann variety, which is constructed
by the Pl\"{u}cker coordinate equations \cite{Emary, Levay1,Hosh4}.
In this geometrical construction one define a measure of
entanglement in terms of Pl\"{u}cker coordinate equations, which is
invariant under action of SLOCC by construction. As an example, let
us consider the quantum system $\mathcal{Q}^{p}_{3}(2,2,2)$. For
such three-qubit states, if for instance the subsystem
$\mathcal{Q}_{1}$ is unentangled with subsystems
$\mathcal{Q}_{2}\mathcal{Q}_{3}$, then we have
\begin{equation}
\mathrm{M}_{1}=\left(%
\begin{array}{cccc}
  \alpha_{1,1,1} & \alpha_{1,1,2}&\alpha_{1,2,1}&\alpha_{1,2,2} \\
 \alpha_{2,1,1} & \alpha_{2,1,2}&\alpha_{2,2,1}&\alpha_{2,2,2} \\
\end{array}%
\right),~\mathrm{M}_{2}=\left(%
\begin{array}{cccc}
  \alpha_{1,1,1} & \alpha_{1,1,2}&\alpha_{2,1,1}&\alpha_{2,1,2} \\
 \alpha_{1,2,1} & \alpha_{1,2,2}&\alpha_{2,2,1}&\alpha_{2,2,2} \\
\end{array}%
\right),
\end{equation}
and $\mathrm{M}_{3}=\left(%
\begin{array}{cccc}
  \alpha_{1,1,1} & \alpha_{2,1,1}&\alpha_{1,2,1}&\alpha_{2,2,1} \\
 \alpha_{1,1,2} & \alpha_{2,1,2}&\alpha_{1,2,2}&\alpha_{2,2,2} \\
\end{array}%
\right)$. To illustrate this construction let us look closer to
the first term of above measure for a three-qubit state
\begin{eqnarray}
&&\Lambda_{1,2}(\mathrm{M}_{1})\overline{\Lambda}_{1,2}(\mathrm{M}_{1})=\\\nonumber&&
|\alpha_{1,1,1}\alpha_{2,1,2}-\alpha_{1,1,2}\alpha_{2,1,1}|^{2}
+|\alpha_{1,1,1}\alpha_{2,2,1}-\alpha_{1,2,1}\alpha_{2,1,1}|^{2}\\\nonumber&&
+|\alpha_{1,1,1}\alpha_{2,2,2}-\alpha_{1,2,2}\alpha_{2,1,1}|^{2}+
|\alpha_{1,1,2}\alpha_{2,2,1}-\alpha_{1,2,1}\alpha_{2,1,2}|^{2}
\\\nonumber&&
+|\alpha_{1,1,2}\alpha_{2,2,2}-\alpha_{1,2,2}\alpha_{2,1,2}|^{2}+
|\alpha_{1,2,1}\alpha_{2,2,2}-\alpha_{1,2,2}\alpha_{2,2,1}|^{2}.
\end{eqnarray}
 $\Lambda_{1,2}(\mathrm{M}_{2})\overline{\Lambda}_{1,2}(\mathrm{M}_{2})$ and
 $\Lambda_{1,2}(\mathrm{M}_{3})\overline{\Lambda}_{1,2}(\mathrm{M}_{3})
 $ can be constructed in a similar ways. Finally, a measure of entanglement for three-qubit states
 is given by
\begin{eqnarray}
\mathcal{E}(\mathcal{Q}^{p}_{3}(2,2,2))&=&\left(\mathcal{N}_{3}\sum^{3}_{j=1}
\Lambda_{1,2}(\mathrm{M}_{j})\overline{\Lambda}_{1,2}(\mathrm{M}_{j})
\right)^{1/2}.
\end{eqnarray}
This measure of entanglement for three-qubit states coincide with
entanglement monotones given in Ref. \cite{Hosh4}.

\section{Entanglement measure for general pure multipartite states}\label{Mpartite}
The generalization of our entanglement measure for multi-qubit
states on general pure multipartite states can be done in a
straightforward manner. Let an operator
$\Lambda_{\mu,\nu}(\mathrm{M}_{j})=\upsilon_{\mu}(\mathrm{M}_{j})\wedge
\upsilon_{\nu}(\mathrm{M}_{j})$ be the wedge product between row
number $\mu$ and $\nu$ of  matrices $\mathrm{M}_{j}$ for all $j$,
which is constructed by coefficient of a general quantum system
$\mathcal{Q}^{p}_{m}(N_{1},\ldots,N_{m})$. For example, we have
\begin{equation}
  \mathrm{M}_{1}=\left(%
\begin{array}{cccc}
  \alpha_{1,1,\ldots,1} & \alpha_{1,1,\ldots,2} &  \ldots & \alpha_{1,N_{2},\ldots,N_{m}}\\
  \alpha_{2,1,\ldots,1} & \alpha_{2,1,\ldots,2} & \ldots & \alpha_{2,N_{2},\ldots,N_{m}}\\
  \vdots & \vdots & \ldots & \vdots\\
  \alpha_{N_{1},1,\ldots,1} & \alpha_{N_{1},1,\ldots,N_{m}} & \ldots & \alpha_{N_{1},N_{2},\ldots,N_{m}}\\
\end{array}%
\right) \\
\end{equation}
$\mathrm{M}_{2}\ldots,\mathrm{M}_{m}$ can be constructed in a
similar ways by permutation of indices as in the case of
multi-qubit states. Then we can define an entanglement measure for
general pure multipartite states by
\begin{eqnarray}
\mathcal{E}(\mathcal{Q}^{p}_{m}(N_{1},\ldots,N_{m}))&=&\left(\mathcal{N}_{m}\sum^{m}_{j=1}\sum_{\forall
\nu>\mu=1}\Lambda_{\mu,\nu}(\mathrm{M}_{j})\overline{\Lambda}_{\mu,\nu}(\mathrm{M}_{j})\right)^{1/2}.
\end{eqnarray}
In this construction, the entanglement measure vanishes on product
states  and it is entanglement monotones. However this result need
further investigation.

\section{Conclusion}
In this paper, we have derived a measure of entanglement that
coincides with concurrence of a general pure bipartite state based
on mapping of a tensor
 product space on an alternating tensor product space defined by a wedge
 product. Moreover, we have constructed a measure of entanglement for
 a pure
 multi-qubit state, which is entanglement monotones. Furthermore,
 we have generalized this construction into a general pure
 multipartite state.

\begin{flushleft}
\textbf{Acknowledgments:} The author acknowledge useful
discussions with Gunnar Bj\"{o}rk. The author also would like to
thank Jan Bogdanski. This work was supported by the Wenner-Gren
Foundations.
\end{flushleft}

\end{document}